\documentclass[a4paper,aps,prl,twocolumn,amsmath,amssymb,showpacs,10pt]{revtex4}
\usepackage{epsfig}
\usepackage{graphicx}
\usepackage{amsmath,amssymb}
\newcommand{\be}{\begin{equation}}
\newcommand{\ee}{\end{equation}}

\newcommand{\zncro}{\text{Zn}_{1-x}\text{Cd}_x\text{Cr}_2\text{O}_4}

\newcommand{\scgo}{\text{SrCr}_{8-x}\text{Ga}_{4+x}\text{O}_{19}}
\newcommand{\ymoo}{\text{Y}_2\text{Mo}_2 \text{O}_7}

\begin{document}
\title{Spin Freezing in Geometrically Frustrated Antiferromagnets with
  Weak Disorder}
\author{T. E. Saunders and J. T. Chalker}
\affiliation{Theoretical Physics, University
  of Oxford, 1 Keble Road, Oxford, OX1 3NP, United Kingdom}
\date{\today}
\begin{abstract}
We investigate the consequences for geometrically frustrated
antiferromagnets of weak disorder in the strength of exchange interactions.
Taking as a model the classical Heisenberg antiferromagnet with nearest
neighbour exchange on the pyrochlore lattice, we examine
low-temperature behaviour. We show that random exchange generates
long-range effective interactions within the extensively degenerate ground states of the clean system.
Using Monte Carlo simulations, we find a spin glass
transition at a temperature set by the disorder strength.
Disorder of this type, which is generated by random strains in the
presence of magnetoelastic coupling, may account for the spin freezing
observed in many geometrically frustrated magnets.
\end{abstract}


\maketitle

Geometrically frustrated magnets -- materials in which magnetic ions
form a lattice consisting of frustrated units such as triangles and
tetrahedra -- characteristically remain in the paramagnetic phase even
at temperatures that are small on the scale set by 
exchange interactions. Nevertheless, at sufficiently low temperature most
examples exhibit spin freezing, with a large value (in the range from 10 to
150) for the ratio $|\Theta_{\rm CW}|/T_{\rm F}$ of the magnitude
of the Curie-Weiss constant to the freezing temperature as an
identifying feature \cite{reviews}.
Indications of freezing include: a difference between
field-cooled and zero-field-cooled susceptibilities \cite{ramirez1990,ramirez1992,martinez1992,wills1998,wills2000a,gingras1997};
a suppression of inelastic magnetic neutron scattering \cite{martinez1992,lee1996,wills1998,gardner1999}; and in
some cases a divergence in the non-linear
susceptibility on approaching $T_{\rm F}$, as in
conventional spin glasses \cite{binder1986}. Some well-studied systems are
$\scgo$ (with a layered structure consisting of slabs
from the pyrochlore lattice) \cite{ramirez1990,ramirez1992,martinez1992,lee1996}, the {\it kagom\'e}
antiferromagnet hydronium jarosite \cite{wills1998,wills2000a}, 
and the pyrochlore $\ymoo$ \cite{gingras1997,gardner1999}.

The origin of this spin freezing has long been puzzling.
On the theoretical side, it is established that freezing is absent
from some simple models
without disorder, including the classical Heisenberg antiferromagnet with nearest
neighbour interactions on the pyrochlore lattice \cite{moessner1998}.
While a more realistic treatment should take account of disorder and
of various residual interactions, in some of these
materials there appears to be little structural disorder \cite{wills1998}, and in
others \cite{ramirez1992} $T_{\rm F}$ is rather insensitive to the identified
form of disorder, dilution at magnetic sites.

In this context, recent experiments which show the importance of
random strains in two pyrochlore materials are particularly interesting, since
via magnetoelastic coupling such strains will lead to local variations
in the strength of exchange interactions. One material is $\ymoo$,
in which disorder in $\text{Mo}{-}\text{Mo}$ distances has been
revealed using XAFS \cite{booth2000}.
The other is $\zncro$.
Unusually for a geometrically frustrated magnet, in pure form
($x{=}0$) this has a first-order transition to a low temperature
phase in which magnetic degeneracy is lifted by a lattice distortion
and there is N\'eel order \cite{lee2000}. 
This phase has a striking sensitivity to substitution of Cd for Zn:
$x{=}0.03$ is sufficient to suppress N\'eel order completely, with spin
freezing taking its place \cite{ratcliff2002}. The effect is argued \cite{ratcliff2002}
to arise from strains around $\text{Cd}$ sites, generated because
$\text{Cd}^{2+}$ has a larger ionic radius than $\text{Zn}^{2+}$.
In remarkable contrast, N\'eel order is much less sensitive
to magnetic dilution by substitution of Gd for Cr: it survives
with up to 25\% of magnetic ions removed \cite{ratcliff2002}.

Taking the background outlined above as motivation, our aim in this
paper is to discuss the effects of weak exchange randomness
in geometrically frustrated antiferromagnets. 
We focus on the classical Heisenberg model with nearest neighbour
interactions on the pyrochlore lattice, 
because of the large body of experimental work on pyrochlore
antiferromagnets, and
because its behaviour
without disorder is well understood \cite{moessner1998}.
In particular, the disorder-free model is known to have an extensively
degenerate, connected ground state manifold \cite{moessner1998}, and a dipolar form for
spin correlations in the limit of low temperature $T$ \cite{isakov2004,henley2005}.
In the presence of exchange disorder one expects two regimes according to its amplitude.
Characterising interactions by their average $J$
and a magnitude $\Delta$ of fluctuations, for strong disorder $(\Delta \gtrsim J)$ the model is
simply a conventional example of a spin glass. Our interest lies
instead with the weak disorder limit $(\Delta \ll J)$ in which exchange
randomness acts as a perturbation lifting the ground-state degeneracy 
of the clean system. 
In the following we show that projection of fluctuations in
nearest-neighbour
exchange interactions
into the ground state manifold of the clean system generates
long-range effective interactions. 
In addition, using Monte Carlo simulations with parallel
tempering, we show for $\Delta \ll J$ that the model has a transition at 
a temperature $T_{\rm F} \propto \Delta$. As the transition is
approached from above, the spin glass susceptibility diverges, and
for $T<T_{\rm F}$ there is long-range spin glass order.

We start from the Hamiltonian 
\be
\label{hamilton}
\mathcal{H} = \sum_{ij} J_{ij} \mathbf{S}_i.\mathbf{S}_j \,,
\ee
in which classical spins $\mathbf{S}_i$ are three-component unit vectors
at the sites $i$ of a pyrochlore lattice, and exchange interactions $J_{ij}$
are non-zero only between neighbouring pairs of sites.
We begin with a qualitative discussion of the special features
of this model at weak disorder.

As a first step, consider a single tetrahedron taken from this
lattice, with spins $\mathbf{S}_1 \ldots \mathbf{S}_4$ at the vertices.
Its ground states are the configurations for which 
$\sum_i \mathbf{S}_i = \mathbf{0}$. With all $J_{ij}$ equal, spin
stiffness is zero in this toy problem in the sense that, within the
set of ground states, the orientations of a pair of spins can be
chosen arbitrarily. The consequences of small variations
in $J_{ij}$ with amplitude $\Delta$ have been set out in Ref.~\cite{tchernyshyov2002}:
generically, a unique ground state is selected (up to global spin
rotations) in which all four spins are collinear and the spin pairs linked by the strongest 
interactions are arranged antiparallel. Variations in $J_{ij}$ hence induce
a ground state stiffness, since changes in the relative
orientation of a pair of spins cost an energy ${\cal O}(\Delta)$.

Moving to the full, pyrochlore lattice problem, we next
argue that weak exchange randomness generates {\it long-range} effective
couplings. The logic is as follows. Without disorder $J_{ij}$ can be block-diagonalised
by Fourier transform. Its spectrum has four branches in the Brillouin
zone. The lowest two branches are degenerate and independent of wavevector, 
mirroring the ground state degeneracy of the model. In the limit
$\Delta \ll J$ it is natural to project the matrix $J_{ij}$ 
onto this degenerate subspace. The matrix elements
$P_{ij}$ of the projection operator have a dipolar form
\cite{isakov2005}:
they decrease as $|\mathbf{r}_i - \mathbf{r}_j|^{-3}$ for large
$|\mathbf{r}_i - \mathbf{r}_j|$. Hence so does the projected
interaction matrix.
In the context of conventional spin glasses, dipolar interactions
have been shown not to be sufficiently long ranged to 
generate different critical behaviour from that with
short-range interactions \cite{bray}; it is not clear at
present
whether this conclusion carries over to the problem
we are concerned with.

The interaction matrix and, in the limit $\Delta \ll J$, its projected version,
enter directly into a calculation in which the fixed spin length
$|\mathbf{S}_i|^2=1$ is treated within the spherical approximation
$\sum_i |\mathbf{S}_i|^2 = N_s$ (where the sum is over $N_s$ spins
in the lattice). In the absence of disorder such an approach is 
exact for a model in which the number of spin components $n\rightarrow
\infty$ \cite{canals2001}, and gives an excellent treatment
of low  $T$ correlations at all $n$ \cite{isakov2004}. Applying it and the replica
method, with a Gaussian distribution for $J_{ij}$ of variance $\Delta^2$,
we find \cite{future} for $\Delta \ll J$ that there is a spin glass transition
at a temperature independent of $J$ and proportional to $\Delta$:
\be
k_{\rm B}T_{\rm F} = \sqrt{{8}/{3}}\,\, \Delta\,. 
\ee

A difficulty in proceeding further with a conventional replica
treatment lies in restricting 
spin configurations to the manifold of ground states 
for the model without disorder,
as is necessary for $\Delta, T \ll J$.
A natural and elegant way of building in that
constraint is to parameterise the ground states in terms of
a gauge field, as set out in Refs.~\cite{isakov2004,henley2005}.
We next examine how the effects of weak exchange disorder
can be introduced into that formulation.
The essence of the gauge field parameterisation 
for the model without disorder can be summarised as follows.
A set of vector fields $\mathbf{B}^a(\mathbf{r})$
is introduced to represent spin configurations,
with one field for each spin component.
The mapping between a spin configuration and vector fields
(see \cite{isakov2004,henley2005} for details) is made 
in such a way that the condition for a configuration to be a ground
state translates into the condition
$\mathbf{\nabla}\cdot \mathbf{B}(\mathbf{r})=0$. After
coarse-graining, the fluxes $\mathbf{B}^a(\mathbf{r})$ are treated as continuous,
divergence-free fields, with a statistical weight $e^{-S_0}$ (before
normalisation) given by
\be
S_0 = \frac{\kappa}{2}\int \text{d}^3\mathbf{r} \sum_a |\mathbf{B}^a(\mathbf{r})|^2\,,
\ee
where the stiffness $\kappa$ is determined by microscopic details of
the model. From the example of a single tetrahedron (and the details
of the mapping between spins and vector fields), we know that
the effect of small variations in $J_{ij}$ is to favour a 
specific axis for flux. This axis turns out to be
one of the cubic crystal axes: the particular one selected 
depends on the values of $J_{ij}$ and varies randomly
from one tetrahedron to another. 
We therefore propose an effective theory for a geometrically
frustrated antiferromagnet with weak exhange disorder, taking 
$S_{\rm eff} = S_0 + S_{\rm dis} + S_{\rm
  int}$
with
\be
S_{\rm dis} = - \beta \Delta \int \text{d}^3\mathbf{r} 
\sum_a \left[\mathbf{B}^a(\mathbf{r})\cdot \mathbf{n}(\mathbf{r})\right]^2
\ee
and 
\be
S_{\rm int} =
\beta u \int \text{d}^3\mathbf{r}  
\big[\sum_a|\mathbf{B}^a(\mathbf{r})|^2\big]^2\,.
\ee
Here, the preferred local axis for flux is defined by the random field
$\mathbf{n}(\mathbf{r})$, which has average
$[{n}_i(\mathbf{r})]_{\rm av} = {0}$ and variance
$[{n}_i(\mathbf{r}){n}_j(\mathbf{r'})]_{\rm av} =
\delta_{ij} \delta(\mathbf{r} - \mathbf{r'})$. 
Inverse temperature is denoted by $\beta$: $S_{\rm dis}$ dominates
over $S_0$ at low temperature, since the first 
is an energetic contribution while the second is entropic.
Microscopics imply an upper bound to $|\mathbf{B}^a(\mathbf{r})|$,
imposed here by $S_{\rm int}$ with phenomenological coefficient $u$.
Note that $S_{\rm eff}$, like ${\cal H}$ is invariant under global spin
rotations, which translate to rotations between the different $\mathbf{B}^a(\mathbf{r})$.
In the langauge of this effective theory, spin freezing is
condensation of flux into a specific arrangement
favoured by disorder. 


To search for such freezing, we turn to Monte Carlo
simulations for the model of Eq.~(\ref{hamilton}). Our work builds
on the initial investigation of Ref.~\cite{bellier-castella2001}, but is much
more detailed. In common with a recent study of the spin glass
transition
in the three-dimensional Edwards-Anderson Heisenberg model
\cite{lee2003}, we use parallel tempering \cite{parallel} to reach
equilibrium at low temperature. Some details are as follows.
We take $J_{ij}$ independently and uniformly distributed on the
interval $[J-\Delta, J+\Delta]$ with $0.05 \leq \Delta/J\leq 0.2$,
and study the temperature range $10^{-2}\leq T/J \leq 1$.
System sizes, specified by the linear dimension $L$ in units of the
lattice constant and by the number of spins
$N_s=4L^3$, are $2\leq L \leq 7$ and $32 \leq N_s \leq 1372$.
Run lengths vary from $5 \times 10^3$ Monte Carlo
steps per spin (MCS) for $L=2$ to $2 \times 10^5$ for $L=7$.
Results are averaged over a number of disorder realisations varying
from $10^3$ for $L=2$ to $200$ for $L=7$.

We present data for three quantities: the heat capacity 
per spin $C_v$, 
the spin glass correlation function $C(\mathbf{r})$, and the
spin glass susceptibility $\chi$. The last two are defined in terms of
the behaviour of two copies of a system with
identical disorder. Labelling the copies with $l=1,2$, 
denoting thermal averages in each copy by $\langle \ldots \rangle_l$
and a disorder average by $[\dots ]_{\mathrm av}$, we have
\be
\label{correl}
C(\mathbf{r}) = [\langle \mathbf{S}(0).\mathbf{S}(\mathbf{r})\rangle_1 
\langle \mathbf{S}(0).\mathbf{S}(\mathbf{r}) \rangle_2]_{\mathrm av}
\ee 
and 
\be
\chi = \sum_{\mathbf{r}} C(\mathbf{r})\,.
\ee

\begin{figure}[tb]
\begin{center}
\includegraphics[width=55mm,angle=270]{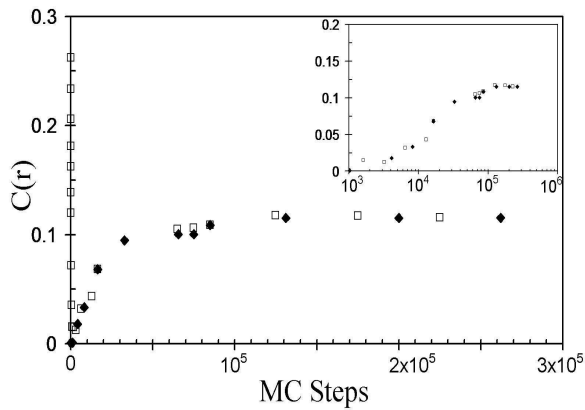}
\caption{
Evolution of $C(\mathbf{r})$ with Monte Carlo time (note logarithmic
scale) for $T/J=0.01$, $\Delta/J = 0.1$, $L=7$, and maximum $r$;
upper and lower curves from different initial states, as described in text. 
}
\label{equilibration}
\end{center}
\vspace{-5mm}
\end{figure}

First, as a test for equilibration, we examine $C(\mathbf{r})$
as a function of simulation time, comparing initial conditions
for which  $C(\mathbf{r})=1$ (both copies initially in the
same N\'eel ground state of the diorder-free model) with ones for
which $C(\mathbf{r})=0$ (each copy initially an independent random
spin configuration). Results are shown in Fig.~\ref{equilibration}
for the most demanding case (maximum $L$ and $r$, minimum $T$).
On this basis, for $L=7$ we collect data after discarding an initial $1\times 10^{5}$ MCS. 
Equilibration is much more rapid at smaller $L$ or larger $T$.

\begin{figure}[t]
\begin{center}
\epsfig {figure=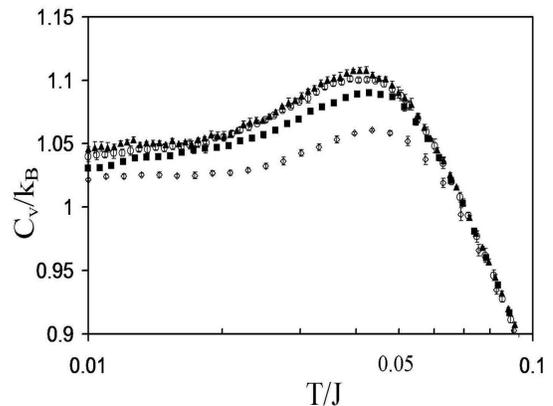, width=55mm,angle=270}
\caption{$C_v$ (in units of $k_{\rm B}$) vs $T/J$ for $\Delta/J=0.1$ and (from top to bottom)
$L{=}6$, 5, 4 and 3.}
\label{heat-capacity}
\end{center}
\vspace{-5mm}
\end{figure}

The heat capacity, shown in Fig.~\ref{heat-capacity}, varies smoothly
with $T$, as expected in the absence of
a N\'eel ordering transition. The consequences of exchange randomness are revealed
in the limiting value of $C_v$ at small $T$: without disorder this
is \cite{moessner1998} $3k_{\rm B}/4$ for large $L$, reflecting
the zero modes ($1/4$ of all degrees of freedom) of the ground states,
but with disorder it rises to $k_{\rm B}$, because all zero modes
are removed, except for the three arising from global spin rotations.

\begin{figure}[t]
\begin{center}
\epsfig {figure=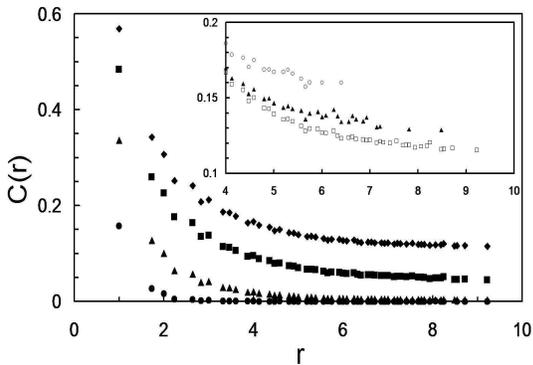, width=50mm,angle=270}
\caption{$C(\mathbf{r})$ vs $r$ for $L=7$ and $\Delta/J = 0.1$
at (from top to bottom) $T/J=0.01$, 0.02, 0.04, and 0.1. Inset:
dependence on system size at $T/J=0.01$, $L=5$, 6, 7.}
\label{corr1}
\end{center}
\vspace{-5mm}
\end{figure}

\begin{figure}[t]
\begin{center}
\epsfig {figure=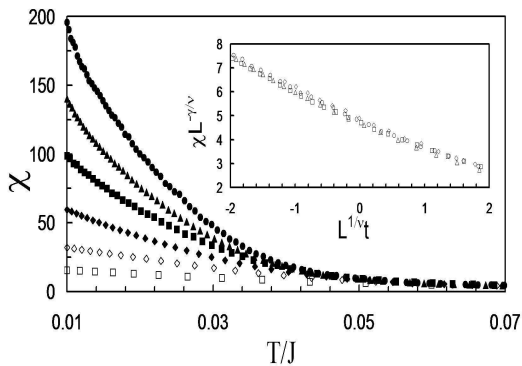, width=50mm,angle=270}
\caption{$\chi$ vs $T/J$ for $\Delta=0.1J$ and system sizes from $L=2$
to  $L=7$. Inset: scaling collapse.}
\label{corrT}
\end{center}
\vspace{-5mm}
\end{figure}

Next, we present the central result of our simulations,
the behaviour of  the spin glass correlation function $C(\mathbf{r})$,
shown in Fig.~\ref{corr1}. We believe this provides clear evidence
that $C(\mathbf{r})$ is non-zero at large $r$ below 
a transition temperature $T_{\rm F}$.
From simple inspection of this figure $0.04<T_{\rm F}<0.02$ at $\Delta/J =
0.1 $. In an effort to determine $T_{\rm F}$ more precisely, 
and to illustrate finite-size effects, we turn to the
susceptibility $\chi$, shown in Fig.~\ref{corrT}. The rapid
increase in $\chi$ with $L$ at low $T$ is clear.
Close to $T_{\rm F}$ one expects the finite-size scaling
behaviour
\be
\chi(T,L) = L^{\gamma/\nu} f(L^{1/\nu} t)
\ee
where $t=(T-T_c)/T_c$, and $\nu$ and $\gamma$ are the standard critical
exponents for the correlation length and susceptibility. 
Scaling collapse of the data is shown in the inset to
Fig.~\ref{corrT}, with $T_{\rm F} = 0.023$, $\nu=1$ and $\gamma=1.45$.
Uncertainties in these parameters are hard to quantify
because finite size effects are large for $L=2$  and $L=3$ ($N_s = 32$ and
$N_s = 108$); omitting these sizes, for $\Delta/J=0.1$ collapse is obtained 
with
$0.020 \leq T_{\rm F} \leq 0.32$, $0.9 \leq \nu \leq 1.2$ and $1\leq \gamma
\leq 1.6$.

\begin{figure}[tb]
\begin{center}
\epsfig {figure=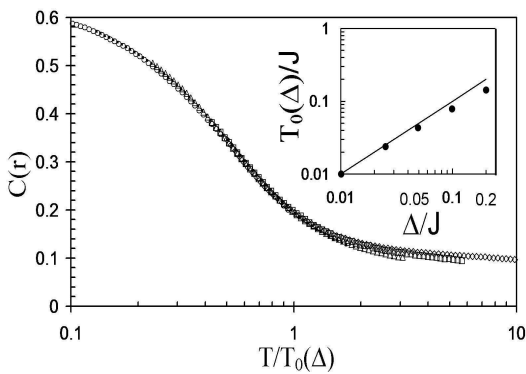, width=50mm,angle=270}
\caption{Dependence on disorder strength $\Delta$: $C(\mathbf{r})$ for
    $r=4$ as a function of $T/T_0(\Delta)$. Inset:
    $T_0(\Delta)/J$ vs $\Delta/J$; line is guide to eye.} 
\label{corr}
\end{center}
\end{figure}

Finally, we examine the dependence of $T_{\rm F}$ on $\Delta$. On
dimensional grounds, one has $T_{\rm F}/\Delta = g(\Delta/J)$, and
from our discussion of spin stiffness in a single tetrahedron
we expect $g(x)$ to be finite in the weak disorder limit $x\rightarrow
0$. We have evaluated $C(\mathbf{r})$ as a function of $T$ for 
$\Delta/J = 0.2$, 0.1, 0.75 and 0.05, considering only
$L=5$ ($N_s=500$) because of limitations on computational resources.
As shown in Fig.~\ref{corr}, the dependence of data for
$C(\mathbf{r})$ at fixed $r$
on $T$ and $\Delta$ can be reduced to a single scaled variable
$T/T_0(\Delta)$,
and $T_0(\Delta) \propto \Delta$ for small $\Delta$.

In summary, we have shown that weak exchange randomness in the
classical Heisenberg antiferromagnet on the pyrochlore lattice
generates long-range effective interactions, and that these
are responsible for a spin glass transition at a temperature
set by the disorder strength. We suggest that this may account
for spin freezing observed in many geometrically frustrated magnets.

We thank C. Broholm, P. C. W. Holdsworth, R. Moessner, and M. A. Moore
for helpful discussions. This work was supported by EPSRC Grant No.
GR/R83712/01. It was completed while JTC was a visitor at KITP Santa
Barbara, supported by NSF Grant No. PHY99-07949.

\end{document}